\documentclass[twocolumn]{aastex631}
\usepackage{graphicx,xspace,apjfonts}

\newcommand\Rocrit{\mathrm{Ro_{crit}}}
\newcommand\Rosun{\mathrm{Ro_\odot}}
\newcommand\xray{\mbox{X-ray}\xspace}

\shorttitle{Weakened Magnetic Braking}
\shortauthors{Metcalfe et al.}

\journalinfo{The Astrophysical Journal Letters (accepted)}

\begin{document}

\title{\Large Weakened Magnetic Braking Signals the Collapse of the Global Stellar Dynamo}

\author[0000-0003-4034-0416]{Travis S.~Metcalfe} 
\affiliation{Center for Solar-Stellar Connections, White Dwarf Research Corporation, 9020 Brumm Trail, Golden, CO 80403, USA}

\author[0000-0002-4284-8638]{Jennifer L.~van~Saders} 
\affiliation{Institute for Astronomy, University of Hawai`i, 2680 Woodlawn Drive, Honolulu, HI 96822, USA}

\author[0000-0002-7549-7766]{Marc H.~Pinsonneault} 
\affiliation{Department of Astronomy, The Ohio State University, 140 West 18th Avenue, Columbus, OH 43210, USA}

\author[0000-0002-1242-5124]{Thomas R.~Ayres} 
\affiliation{Center for Astrophysics and Space Astronomy, 389 UCB, University of Colorado, Boulder, CO 80309, USA}

\author[0000-0003-3061-4591]{Oleg Kochukhov} 
\affiliation{Department of Physics and Astronomy, Uppsala University, Box 516, SE-75120 Uppsala, Sweden}

\author[0000-0002-3481-9052]{Keivan G.~Stassun} 
\affiliation{Vanderbilt University, Department of Physics \& Astronomy, 6301 Stevenson Center Lane, Nashville, TN 37235, USA}

\author[0000-0002-3020-9409]{Adam J.~Finley} 
\affiliation{Universit\'e Paris-Saclay, Universit\'e Paris Cit\'e, CEA, CNRS, AIM, F-91191, Gif-sur-Yvette, France}

\author[0000-0001-5986-3423]{Victor See} 
\affiliation{School of Physics \& Astronomy, University of Birmingham, Edgbaston, Birmingham B15 2TT, UK}

\author[0000-0002-0551-046X]{Ilya V.~Ilyin} 
\affiliation{Leibniz-Institut f\"ur Astrophysik Potsdam (AIP), An der Sternwarte 16, D-14482 Potsdam, Germany}

\author[0000-0002-6192-6494]{Klaus G.~Strassmeier} 
\affiliation{Leibniz-Institut f\"ur Astrophysik Potsdam (AIP), An der Sternwarte 16, D-14482 Potsdam, Germany}

\begin{abstract}

Weakened magnetic braking (WMB) was originally proposed in 2016 to explain anomalously 
rapid rotation in old field stars observed by the Kepler mission. The proximate cause was 
suggested to be a transition in magnetic morphology from larger to smaller spatial 
scales. In a series of papers over the past five years, we have collected 
spectropolarimetric measurements to constrain the large-scale magnetic fields for a 
sample of stars spanning this transition, including a range of spectral types from late F 
to early K. During this time, we gradually improved our methods for estimating the wind 
braking torque in each of our targets, and for evaluating the associated uncertainties. 
Here, we reanalyze the entire sample with a focus on uniformity for the relevant 
observational inputs. We supplement the sample with two additional active stars to 
provide more context for the evolution of wind braking torque with stellar Rossby number 
(Ro). The results demonstrate unambiguously that standard spin-down models can reproduce 
the evolution of wind braking torque for active stars, but WMB is required to explain the 
subsequent abrupt decrease in torque as Ro approaches a critical value for dynamo 
excitation. This transition is seen in both the large-scale magnetic field and the \xray 
luminosity, indicating weakened coronal heating. We interpret these transitions as 
evidence of a rotational threshold for the influence of Coriolis forces on global 
convective patterns and the resulting inefficiency of the global stellar dynamo.

\end{abstract}


\section{Introduction}\label{sec1}

In astronomy we are fond of dividing stars into classes. Some categories are even useful, 
and the broad principles behind them are tied to fundamental astrophysics. Rotation, and 
its time evolution, is one such case. One of the most fundamental divisions in stellar 
astrophysics---between high and low mass stars---is clearly seen in the evolution of 
their rotation rates \citep[e.g., see][their Fig.1]{Kraft1967}. The median high 
mass star rotates much faster than the median low mass star. High mass stars, as a rule, 
do not spin down due to magnetized winds. Low mass stars do.  Exceptions to these rules 
are important, because they test our assumptions and lead to deeper physical 
understanding. This paper is about weakened magnetic braking (WMB) in low mass stars, and 
its consequences for the origin and evolution of stellar magnetism, and for rotation as 
an age diagnostic.

The standard model of stellar spin-down is a natural starting point because magnetized 
winds are efficient engines for angular momentum loss. The physical model is 
straightforward: the wind co-rotates with the star to a characteristic Alfv\'en radius, 
which is much larger than the radius of the star. \cite{WeberDavis1967} demonstrated that 
the solar wind could remove enough angular momentum to explain the slow solar rotation. 
However, extrapolating beyond the Sun was difficult. Both of the key ingredients---the 
large-scale magnetic field strength and the mass-loss rate---are extremely 
difficult to measure in other stars, and this was even more true in the 1960s than it is 
today.  The next key development in the field was the careful collation of empirical data 
by \cite{Skumanich1972}. He demonstrated that activity diagnostics, the surface lithium 
abundance, and the rotation rate of stars all scale with age as $t^{-1/2}.$ Lithium 
appears here because it can be destroyed by rotationally induced mixing 
\citep{Pinsonneault1997}, and the destruction rate is expected to be proportional to the 
rotation rate. When coupled with the Weber-Davis model, the Skumanich relations can be 
explained if the magnetic field strength is proportional to the rotation rate.

By the 1990s there was a wave of new data. Efficient spectrographs with CCDs could 
measure rotational line broadening, which uncovered a population of young cluster 
stars with a wide range of rotation rates \citep{Soderblom1983, Stauffer1987}. Time 
domain surveys yielded rotation periods from spot modulation, beginning with the 
pioneering Mount Wilson survey \citep{Baliunas1996} and continuing to studies of rotation 
in open clusters, such as the Hyades \citep{Radick1987}. \xray data from ROSAT and 
ultraviolet data from IUE added a wealth of knowledge about coronal and chromospheric 
activity. This was complemented by the development of theoretical models for angular 
momentum evolution \citep{Pinsonneault1989, Pinsonneault1990}. By the end of the 1990s, a 
coherent model for angular momentum evolution had been developed 
\citep{Krishnamurthi1997}. The range of initial rotation rates arose from star-disk 
interactions \citep{Konigl1991, Shu1993}. The magnetic field strength increased with 
rotation rate, as inferred by Skumanich, until it saturated at a critical level 
\citep{MacGregor1991}. A transient phase of core-envelope decoupling, with a timescale of 
tens of Myr, was needed for solar-type stars \citep{Charbonneau1993}.

Time domain surveys from space then arose, and they utterly transformed stellar 
astrophysics. The Kepler mission \citep{Borucki2010} yielded tens of thousands of 
rotation periods for field stars \citep{McQuillan2014}. Rotation was now a viable 
chronometer for large stellar samples, giving rise to the field of gyrochronology 
\citep{Barnes2007}. The limitations of purely solar-scaled models also became apparent. 
In response, there was a burst of new work on magnetized winds \citep{Matt2012, 
Reiners2012, Gallet2013, vanSaders2013, Garraffo2018, Spada2020}. A central insight was 
the governing role of the Rossby number (Ro), or the ratio between rotation period and 
convective overturn timescale. In these models, both the magnetic field strength and the 
mass-loss rates were explicitly tied to Ro, and the results were much more successful at 
explaining the underlying mass trends. However, the new generation of models all used 
indirect proxies for the magnetic fields and the mass loss, scaling them relative to 
global stellar properties.

In stubborn contradiction to theory, stars less active than the Sun failed to slow down 
at the expected rate \citep{vanSaders2016, vanSaders2019}. Furthermore, this weakened 
braking also appeared to be tied to Ro; it occurs at a shorter rotation period for F 
stars than for G, and the threshold for K stars is slower still. Motivated by these 
results, we have engaged in a systematic campaign to bypass activity proxies and more 
directly measure large-scale magnetic field strengths \citep{Metcalfe2021, Metcalfe2022, 
Metcalfe2023a, Metcalfe2024a, Metcalfe2025}. We are also using \xray measurements, as 
opposed to abstract scalings with global properties, to infer mass-loss rates. Finally, 
we have complemented the new data with a new generation of theoretical torque 
calculations \citep{FinleyMatt2018}.

In this paper we engage in a systematic reanalysis, placing the measurements and models 
on a common scale. With the new data and modeling, we are now in a position to address 
several key questions. Is Ro a unique predictor for the onset of weakened braking? Is the 
transition instantaneous or gradual? To what extent does a change in field morphology, 
rather than decreasing field strength, matter? Is there evidence for an anomaly in the 
magnetic field strength, the mass-loss rate, or both? We summarize our uniform sample of 
stellar properties in Section~\ref{sec2}, we describe our homogeneous approach to stellar 
modeling in Section~\ref{sec3}, and we analyze and discuss the results in 
Section~\ref{sec4}.

\section{Observations}\label{sec2}

In our previous work, we adopted stellar properties from a variety of sources in an 
effort to make the resulting estimates of wind braking torque as reliable as possible.  
For this paper, we shift the focus to uniformity across the sample to ensure that our 
inferences of the onset and magnitude of WMB are robust. In this section we describe our 
adopted sources for global stellar properties (\S\ref{sec2.1}), as well as our updated 
approach to constrain the magnetic morphology (\S\ref{sec2.2}) and the mass-loss rate 
(\S\ref{sec2.3}) for each of our targets.

\begin{deluxetable*}{@{}lcCCCCCCCCC@{}}
 \setlength{\tabcolsep}{6pt}
 \tabletypesize{\footnotesize}
 \tablecaption{Uniform Stellar Properties for the Magnetic Braking Sample\label{tab1}}
 \tablehead{\colhead{Star} & \colhead{Name} & \colhead{$|B_{\rm d}|$ (G)} & \colhead{$|B_{\rm q}|$ (G)} & \colhead{$|B_{\rm o}|$ (G)} & \colhead{$\dot{M}$ ($\dot{M}_\odot$)} &
            \colhead{$P_{\rm rot}$ (days)} & \colhead{$M$ ($M_\odot$)} & \colhead{$R$ ($R_\odot$)} & \colhead{Torque ($10^{30}$ erg)} & \colhead{Ro ($\Rosun$)} }
 \startdata
Sun        & \nodata         & 1.54                 & 1.07    & 2.74    & 1.00                 & 25.4                & 1.00        & 1.000         & 0.351^{+0.249}_{-0.151} & 1.000                   \\
HD\,10476  & 107\,Psc        & 4.24                 & 2.77    & 1.37    & 0.64^{+0.76}_{-0.46} & 35\pm0.5            & 0.86\pm0.05 & 0.811\pm0.017 & 0.311^{+0.211}_{-0.169} & 0.935^{+0.073}_{-0.064} \\
HD\,10700  & $\tau$\,Cet     & 0.86^{+0.34}_{-0.34} & \nodata & \nodata & 0.10^{+0.13}_{-0.08} & 34\pm0.5            & 0.82\pm0.05 & 0.836\pm0.020 & 0.030^{+0.041}_{-0.023} & 1.054^{+0.093}_{-0.081} \\
HD\,17051  & $\iota$\,Hor    & 2.13                 & 3.89    & 3.91    & 89.5^{+144}_{-66.5}  & 7.7^{+0.18}_{-0.67} & 1.19\pm0.07 & 1.161\pm0.017 & 51.52^{+72.96}_{-33.75} & 0.562^{+0.162}_{-0.123} \\
HD\,20630  & $\kappa^1$\,Cet & 16.0                 & 15.6    & 11.1    & 118.^{+198}_{-88.9}  & 9\pm0.5             & 1.05\pm0.06 & 0.914\pm0.014 & 96.01^{+88.30}_{-55.94} & 0.339^{+0.043}_{-0.034} \\
HD\,22049  & $\epsilon$\,Eri & 14.6                 & 8.78    & 5.90    & 22.4^{+34.4}_{-16.3} & 12\pm0.5            & 0.86\pm0.05 & 0.694\pm0.014 & 11.85^{+10.25}_{-6.646} & 0.303^{+0.025}_{-0.023} \\
HD\,76151  & \nodata         & 5.98                 & 2.15    & 0.28    & 29.2^{+39.4}_{-19.8} & 15\pm0.5            & 1.05\pm0.06 & 0.964\pm0.018 & 12.98^{+9.933}_{-6.654} & 0.613^{+0.087}_{-0.068} \\
HD\,100180 & 88\,Leo         & 4.03^{+0.53}_{-0.00} & \nodata & \nodata & 6.44^{+8.68}_{-4.49} & 14\pm0.5            & 1.12\pm0.07 & 1.132\pm0.032 & 6.969^{+7.269}_{-3.795} & 0.887^{+0.265}_{-0.184} \\
HD\,101501 & 61\,UMa         & 11.5                 & 12.0    & 6.12    & 29.8^{+36.3}_{-19.0} & 17\pm0.5            & 0.97\pm0.06 & 0.855\pm0.014 & 14.71^{+10.22}_{-6.992} & 0.542^{+0.052}_{-0.046} \\
HD\,103095 & \nodata         & 0.61^{+0.03}_{-0.04} & \nodata & \nodata & 0.05^{+0.07}_{-0.04} & 31\pm0.5            & 0.60\pm0.04 & 0.641\pm0.016 & 0.008^{+0.007}_{-0.005} & 0.933^{+0.068}_{-0.062} \\
HD\,143761 & $\rho$\,CrB     & 1.28^{+0.46}_{-0.46} & \nodata & \nodata & 0.24^{+0.32}_{-0.20} & 17\pm0.5            & 1.05\pm0.06 & 1.300\pm0.025 & 0.529^{+0.701}_{-0.409} & 0.965^{+0.314}_{-0.210} \\
HD\,146233 & 18\,Sco         & 1.34                 & 2.01    & 0.86    & 0.36^{+0.39}_{-0.23} & 22.7\pm0.5          & 1.07\pm0.06 & 1.009\pm0.019 & 0.231^{+0.147}_{-0.110} & 0.985^{+0.172}_{-0.121} \\
HD\,166620 & \nodata         & 2.81^{+0.95}_{-0.94} & \nodata & \nodata & 0.53^{+0.58}_{-0.34} & 43\pm0.5            & 0.83\pm0.05 & 0.771\pm0.019 & 0.135^{+0.156}_{-0.087} & 1.033^{+0.071}_{-0.065} \\
HD\,185144 & $\sigma$\,Dra   & 5.68                 & 4.82    & 4.76    & 4.17^{+5.91}_{-3.01} & 27\pm0.5            & 0.86\pm0.05 & 0.769\pm0.013 & 1.232^{+0.932}_{-0.666} & 0.762^{+0.060}_{-0.052} \\
HD\,186408 & 16\,Cyg\,A      & 0.46^{+0.45}_{-0.44} & \nodata & \nodata & 1.30^{+1.37}_{-0.78} & 20.5^{+2.0}_{-1.1}  & 1.10\pm0.07 & 1.231\pm0.024 & 0.358^{+0.768}_{-0.327} & 1.036^{+0.308}_{-0.215} \\
HD\,186427 & 16\,Cyg\,B      & 0.88^{+1.04}_{-0.73} & \nodata & \nodata & 0.42^{+0.46}_{-0.26} & 21.2^{+1.8}_{-1.5}  & 1.05\pm0.06 & 1.157\pm0.019 & 0.283^{+0.708}_{-0.254} & 0.919^{+0.246}_{-0.157} \\
HD\,217014 & 51\,Peg         & 0.77                 & 0.44    & 0.65    & 0.20^{+0.27}_{-0.17} & 21.9\pm0.4          & 1.10\pm0.07 & 1.174\pm0.023 & 0.168^{+0.125}_{-0.113} & 1.073^{+0.290}_{-0.196} \\
HD\,219134 & \nodata         & 2.39                 & 4.05    & 1.19    & 0.31^{+0.31}_{-0.16} & 42.2\pm0.9          & 0.80\pm0.05 & 0.724\pm0.014 & 0.073^{+0.044}_{-0.028} & 0.892^{+0.058}_{-0.057} \\
 \enddata
 \tablecomments{A machine-readable version of this table includes additional columns ($T_{\rm eff}, \log g, \mathrm{[M/H]}, v \sin i, \log R'_{\rm HK}, P_{\rm cyc}, L_{\rm X}, L_{\rm bol}$) that are not displayed here.}
\end{deluxetable*}
\vspace*{-24pt}

\subsection{Stellar Properties}\label{sec2.1} 

We began by adopting a uniform set of spectroscopic parameters from 
\cite{ValentiFischer2005}, including the effective temperature $T_{\rm eff}$, surface 
gravity $\log g$, metallicity [M/H], and projected rotational velocity $v \sin i$. All of 
our targets were included in this catalog, while the more recent catalog of 
\cite{Brewer2016} contains only a subset.

We used these parameters to obtain empirical constraints on the stellar luminosities and 
radii from an analysis of the broadband spectral energy distribution (SED) for each star, 
following the approach described by \cite{Stassun2016} and \cite{Stassun2017, 
Stassun2018}. This analysis relies on FUV and NUV magnitudes from GALEX, $UBV$ magnitudes 
from \cite{Mermilliod2006}, Str\"omgren $ubvy$ magnitudes from \cite{Paunzen2015}, 
$JHK_S$ magnitudes from 2MASS, and W1--W4 magnitudes from WISE, in some cases spanning 
the full stellar SED from 0.2--20~$\mu$m. We fit the available data for each target using 
Kurucz stellar atmosphere models with the adopted spectroscopic parameters and the 
extinction $A_V$ fixed at zero for these nearby stars. We integrated the resulting model 
SED to determine the bolometric flux at Earth, and we combined this with the Gaia DR3 
parallax \citep{Gaia2021} to calculate the bolometric luminosity $L_{\rm bol}$. The 
values of $L_{\rm bol}$ and $T_{\rm eff}$ yield the stellar radius $R$ from the 
Stefan-Boltzmann relation, while the stellar mass $M$ was derived from the spectroscopic 
parameters using the eclipsing-binary based empirical relations of \cite{Torres2010}.

There is no single source that includes rotation periods $P_{\rm rot}$ for all of our 
targets, but the largest uniform compilation comes from the Mount Wilson survey 
\citep{Baliunas1996, Simpson2010}. We adopted these values for 12 of our 17 targets, with 
the rest derived from Zeeman-Doppler imaging \citep[ZDI;][]{Petit2008, AlvaradoGomez2018, 
Folsom2018} or asteroseismology \citep{Hall2021}. The uniform properties for the sample 
are listed in Table~\ref{tab1}.

\subsection{Magnetic Morphology}\label{sec2.2} 

For the ten targets with existing ZDI maps, we followed the procedures described by 
\cite{Metcalfe2024a} to derive equivalent polar field strengths for the dipole, 
quadrupole, and octupole components of the large-scale magnetic field ($B_{\rm d}$, 
$B_{\rm q}$, $B_{\rm o}$). We adopted these values without revision for 18\,Sco 
\citep{Petit2008, Metcalfe2022}, 61\,UMa \citep{See2019, Metcalfe2023a}, 51\,Peg 
\citep{Metcalfe2024a}, as well as $\epsilon$\,Eri, $\sigma$\,Dra, 107\,Psc, and 
HD\,219134 \citep{Jeffers2014, Folsom2018, Metcalfe2025}. For HD\,76151 we analyzed a new 
ZDI map obtained in 2024 January near its mean activity level \citep{Bellotti2025}, which 
showed better agreement with the rotation period determined by \cite{Baliunas1996}. To 
provide additional context, we also analyzed ZDI maps for the active stars $\iota$\,Hor 
\citep{AlvaradoGomez2018} and $\kappa^1$\,Cet \citep{doNascimento2016}.

For the seven targets with circular polarization (Stokes~$V$) profiles obtained at a 
single rotational phase, we made the conservative assumption that all of the field was in 
the dipole component---maximizing the resulting torque estimate. Following the procedures 
described by \cite{Metcalfe2022}, we modeled each Stokes~$V$ profile with an axisymmetric 
dipole field assuming a fixed stellar inclination. For $\tau$\,Cet the inclination was 
fixed from the orientation of its debris disk \citep{Lawler2014}, while for 16\,Cyg\,A 
and B the inclinations were fixed at the asteroseismic values derived by \cite{Hall2021}. 
For the remaining targets we used the adopted stellar properties ($v\sin i$, $P_{\rm 
rot}$, $R$) to calculate a posterior distribution for the inclination following 
\cite{Bowler2023}, and we adopted the median value. In one case (88\,Leo) the observed 
Stokes~$V$ profile was clearly non-axisymmetric, so we modeled it with a titled dipole 
following the procedures described by \cite{Santos2025}.

\subsection{Mass-Loss Rate}\label{sec2.3} 

We followed the procedures described by \cite{Ayres2025} to obtain a uniform set of \xray 
luminosities $L_{\rm X}$, and we adopted the empirical relation of \cite{Wood2018} to 
estimate mass-loss rates $\dot{M}$ from the resulting \xray surface fluxes $F_{\rm X}$. 
Our approach to determine an \xray luminosity for each target involved reconciling all of 
the available measurements from ROSAT, Chandra, and XMM with the adopted stellar 
properties. Count rates from each of these missions were converted to \xray fluxes at 
Earth using an optimization scheme based on a grid of coronal emission-measure models, 
including a model-based determination of optimum energy conversion factors for each 
instrument. We adopted the mean from the available \xray measurements to obtain a 
representative value of $L_{\rm X}$, with the standard deviation serving as a proxy of 
the long-term variability from stellar cycles and the systematic differences between the 
various instruments.

To estimate mass-loss rates we have previously adopted the empirical relation of 
\cite{Wood2021} for GKM dwarfs, which covers a broader range of $F_{\rm X}$ and has a 
shallower dependence $\dot{M} \propto F_{\rm X}^{0.77\pm0.04}$. This is the more 
conservative choice because it predicts a slower decline in $\dot{M}$ at low $F_{\rm X}$ 
near the onset of WMB. However, the scatter in the \cite{Wood2021} relation is large, 
particularly at high activity levels.  By contrast, the scatter in the \cite{Wood2018} 
relation for GK dwarfs is roughly plus or minus a factor of two, which is consistent with 
the systematic noise floor estimated by \cite{Wood2005}. Having previously demonstrated 
that our conclusions do not depend on this choice, we adopted the steeper relation 
$\dot{M} \propto F_{\rm X}^{1.29\pm0.16}$ from \cite{Wood2018}. We combined the quoted 
factor of two systematic error in quadrature with the errors on $L_{\rm X}$, $R$, and the 
power law exponent to determine the total uncertainty in our estimated mass-loss rates.

\section{Modeling}\label{sec3}

The ultimate goal of our analysis is to determine empirically how the wind braking torque 
changes with the Rossby number across the transition to WMB. In this section we define a 
homogeneous Ro scale from detailed evolutionary modeling (\S\ref{sec3.1}) and we update 
our estimates of the wind braking torque (\S\ref{sec3.2}), drawing from the uniform 
observational inputs described in Section~\ref{sec2}.

\subsection{Rossby Scale}\label{sec3.1}

We defined a uniform Ro scale using the model grids described by \cite{Metcalfe2025}, 
extended to cover the range of masses, metallicities, and evolutionary states in our 
sample. We inferred convective overturn timescales one pressure scale height above the 
convective boundary using a fitting method identical to that in \cite{Metcalfe2025} with 
one difference: we did not utilize $P_{\rm rot}$ as a constraint for the fit, instead 
using only the observational constraints on $T_{\rm eff}$, $R$, and [M/H]. This makes our 
analysis independent of the choice of rotational evolution model at the expense of 
slightly larger uncertainties in the inferred convective overturn timescales, $\tau_{\rm 
c}$. We define the Rossby number as $\textrm{Ro} \equiv P_{\rm rot}/\tau_{\rm c}$ using 
the measured rotation period and the model-inferred overturn timescale with their 
respective uncertainties.

The largest Rossby numbers in our sample are less than $1.1\ \Rosun$, in line with 
expectations from the WMB scenario, despite the fact that the Ro scale defined here is 
agnostic to the rotational evolution model. The WMB hypothesis predicts very mild 
evolution of Ro after the onset of weakened braking, due to an increasing moment of 
inertia under approximate conservation of angular momentum. Using the braking models 
described by \cite{Metcalfe2025}, a solar model with WMB has a main-sequence turnoff 
(core H fraction $<0.0001$) at $\mathrm{Ro}\sim 1.1\ \Rosun$, while standard spin-down 
models would predict $\mathrm{Ro}\sim 1.8\ \Rosun$ at the turnoff.

The large uncertainties on Ro, for the F-type stars in particular, are a reflection of 
the basic behavior of $\tau_{\rm c}$ as a function of surface temperature. F-type stars 
have rapidly thinning convection zones with increasing surface temperature 
\citep{Kraft1967}, and thus rapidly decreasing $\tau_{\rm c}$. Typical observational 
uncertainties on $T_{\rm eff}$ therefore translate into larger uncertainties on Ro 
compared to the cooler stars.

 \begin{figure}[!t]
 \centering\includegraphics[width=\columnwidth]{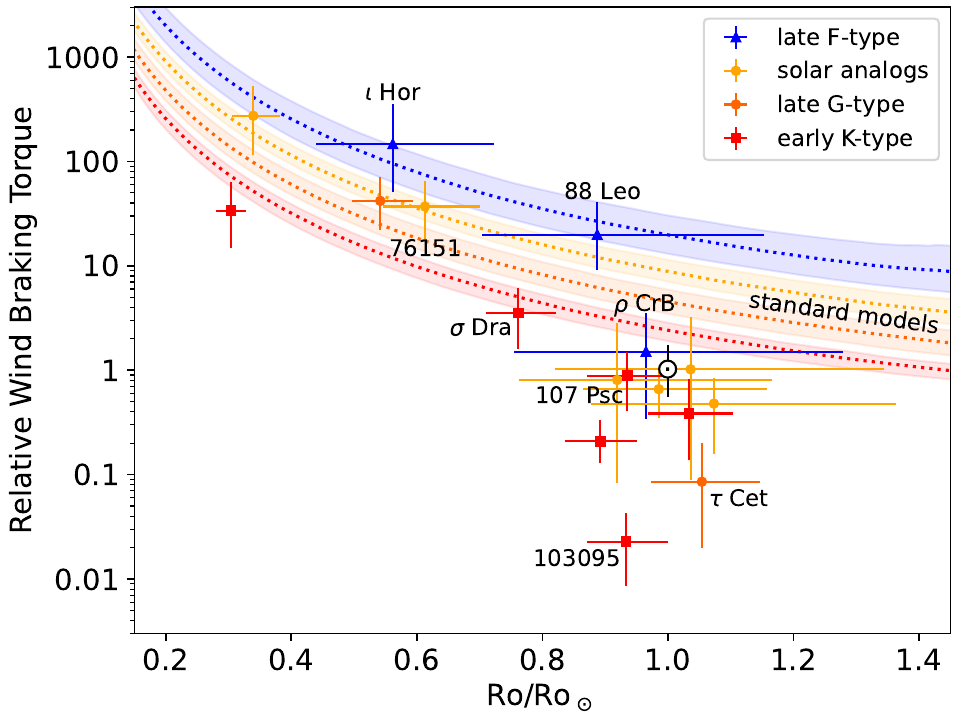} 
 \caption{Relative wind braking torque as a function of Rossby number normalized to the 
solar value. Points are grouped by spectral type as shown in the legend. Standard 
spin-down models for the mean stellar properties within each spectral type are shown for 
reference. The solar point $\odot$ is from \cite{Finley2018}.\label{fig1}}
 \end{figure}

\subsection{Wind Braking Torque}\label{sec3.2}

Adopting the uniform observational inputs described in Section~\ref{sec2}, we updated the 
torque estimates for our sample using the wind braking prescription of 
\cite{FinleyMatt2018}\footnote{\url{https://github.com/travismetcalfe/FinleyMatt2018}}. 
The observational inputs for each star are listed in Table~\ref{tab1}, including the 
large-scale magnetic field components from ZDI maps or Stokes~$V$ snapshots, mass-loss 
rates from the \cite{Wood2018} empirical relation, rotation periods primarily from the 
Mount Wilson survey \citep{Baliunas1996}, stellar masses from the \cite{Torres2010} 
empirical relation, and stellar radii from SED fitting. Uncertainties were determined by 
simultaneously shifting all of the inputs to their $\pm 1\sigma$ values to minimize or 
maximize the torque. The results are illustrated in Figure~\ref{fig1}, with the wind 
braking torque plotted against the Ro scale described in Section~\ref{sec3.1}. Our 
standard spin-down models are shown for the mean stellar properties within each spectral 
type, revealing the mass-dependence of the absolute torque and emphasizing deviations 
from the predicted evolution as Ro approaches and exceeds the solar value.

The results are qualitatively similar to our previously published analyses, showing an 
abrupt change in the estimated wind braking torque as the Rossby number approaches a 
critical value. The inclusion of $\iota$\,Hor to provide additional context for our 
observations of late F-type stars reveals that 88\,Leo already exhibits some evidence of 
WMB, while $\rho$\,CrB remains clearly in the WMB regime with a torque that is 15 times 
weaker than a standard spin-down model. The adoption of a different rotation period for 
HD\,76151 shifts it well below the onset of WMB, while the more slowly rotating solar 
analogs have torques that are \mbox{20--30} times weaker than standard models. The 
results for late G-type stars do not strongly constrain the onset of WMB, but the 
estimated torque for metal-poor $\tau$\,Cet is nearly a factor of 40 below a standard 
model. The K-type star $\sigma$\,Dra appears to be approaching the WMB regime, while the 
more slowly rotating K dwarfs have torques that are \mbox{4--70} times weaker than 
expected from standard models---from the transitional star 107\,Psc to the extremely 
metal-poor star HD\,103095.

 \begin{figure}[!t]
 \centering\includegraphics[width=\columnwidth]{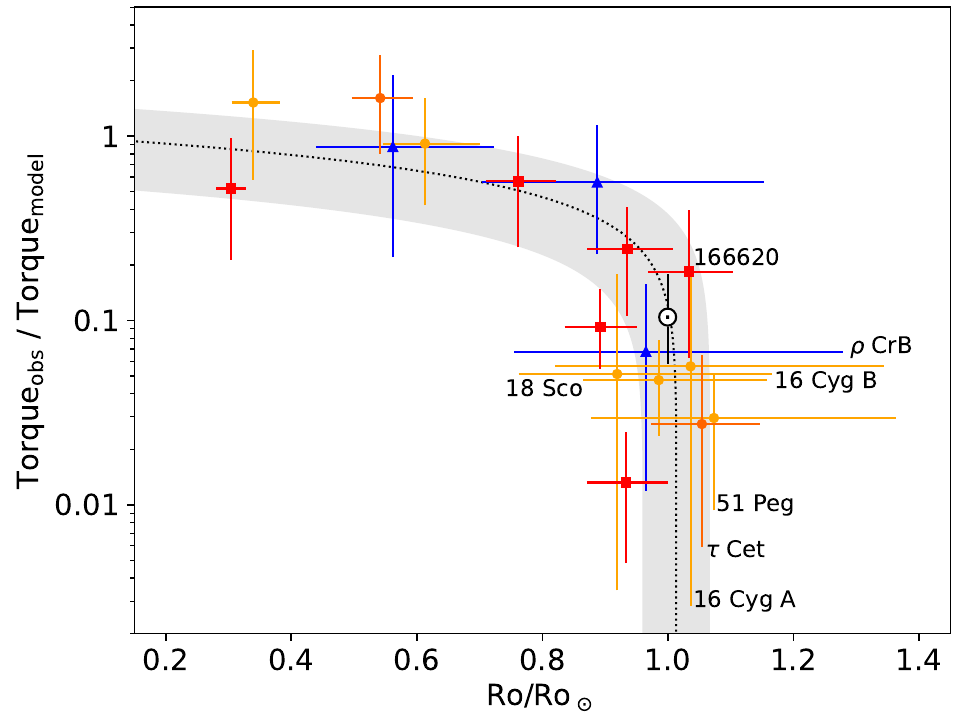} 
 \caption{Ratio of the observationally estimated and model wind braking torque as a 
function of Rossby number normalized to the solar value. The dotted line illustrates the 
fit described in Section~\ref{sec4}, and the 95\% confidence interval is shown as a gray 
shaded region. Flat activity stars and other targets discussed in the text are 
labeled.\label{fig2}}
 \end{figure}

In Figure~\ref{fig2} we show the ratio of the observationally estimated and model torques 
for each of the stars in our sample, normalized to place the observations and models on 
the same scale. This representation allows us to determine the magnitude of the deviation 
of observations from our standard spin-down models, and assess whether the onset of WMB 
is instantaneous or gradual. By construction, the active targets at low Ro are scattered 
around unity, indicating agreement with standard spin-down models. By contrast, the 
targets at higher Ro exhibit an abrupt decrease of nearly two orders of magnitude as Ro 
becomes comparable to the solar value. In the following section, we discuss our 
motivation for fitting the specific functional form that is shown in Figure~\ref{fig2}.

\section{Discussion}\label{sec4}

Our uniform analysis of a sample of targets spanning the transition to WMB motivates a 
paradigm shift in our interpretation of the results. For each of the evolutionary 
sequences that we have previously analyzed, we sought to identify a critical Rossby 
number ($\Rocrit$) beyond which magnetic braking effectively ceased. As we gradually 
expanded the sample, we attempted to trace the onset of WMB back to its root 
causes---from its influence on stellar rotation periods, to a hypothesized shift in 
magnetic morphology, and ultimately to the evolution of the global stellar dynamo. In an 
effort to understand solar cycle variability, \cite{Cameron2017} proposed that the global 
solar dynamo can be considered a ``weakly nonlinear system in the vicinity of a 
supercritical Hopf bifurcation.'' Within this framework, the control parameter is the 
dynamo number $D \sim \mathrm{Ro}^{-2}$ \citep{DurneyLatour1978}, and the system exhibits 
a periodic solution as long as $D$ exceeds a critical value. Below the critical value of 
$D$ (when Ro is sufficiently large), the global dynamo is not excited. Generically, this 
suggests that the solar dynamo should be mildly supercritical \citep{Wavhal2025} and that 
related observables should vary with $N\sqrt{\Rocrit - \mathrm{Ro}}$, where $N$ is a 
normalization constant. The adoption of this functional form suggests that the onset of 
WMB occurs gradually as Ro approaches (rather than exceeds) $\Rocrit$. This appears to be 
supported by the data in Figure~\ref{fig2}, where $N$ is primarily constrained by the 
most active stars in the sample and $\Rocrit$ is largely determined by stars with 
$\mathrm{Ro} \sim \Rosun$.

 \begin{figure*}[!t]
 \includegraphics[width=\columnwidth]{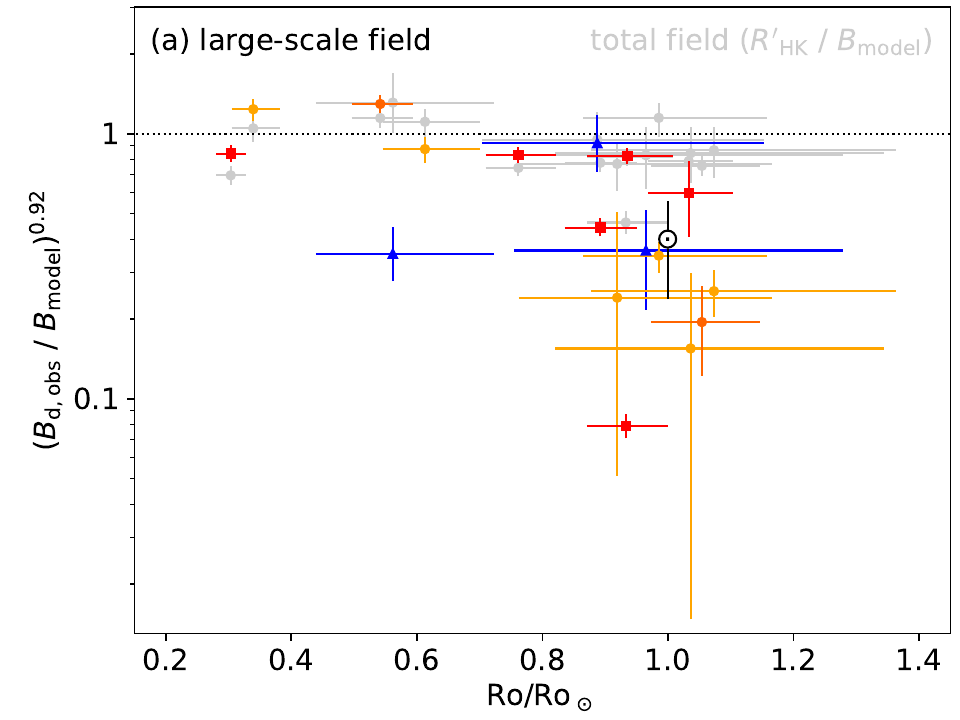}\hfill\includegraphics[width=\columnwidth]{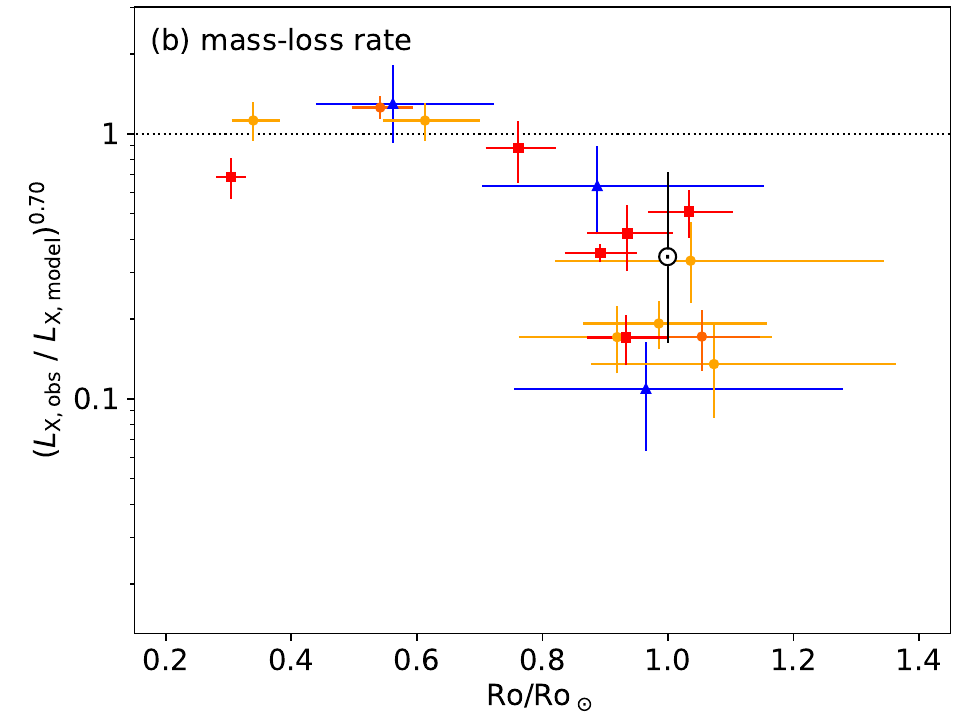}
 \caption{Ratio of observations and model predictions as a function of Rossby number 
normalized to the solar value. Each ratio has been raised to a power that reflects its 
relative contribution to the change in wind braking torque and normalized so that unity 
represents agreement. (a)~Dipole magnetic field strength as a proxy for the large-scale 
field, and (b)~\xray luminosity, reflecting changes in the stellar mass-loss rate. The 
solar dipole field strength is from \cite{Finley2018}, while the \xray luminosity over a 
complete solar cycle is from \cite{Judge2003}.\label{fig3}}
 \end{figure*}

The two axes of Figure~\ref{fig2} are not entirely independent, so we used an iterative 
strategy to estimate the values of $N$ and $\Rocrit$. We initially set the value of 
$N=0.329$ from the mean of the seven most active stars, which agree with our standard 
spin-down models. We then used the stars with $\mathrm{Ro} > 0.7$ to determine the 
optimal value of $\Rocrit=1.014 \pm 0.026\ \Rosun$, considering only the horizontal 
uncertainties. Finally, we fixed $\Rocrit$ at its optimal value and reoptimized the value 
of $N=0.354 \pm 0.077$, considering only the vertical uncertainties. The optimal value of 
$N$ was also within this range when fixing $\Rocrit$ at its $\pm1\sigma$ limits. Although 
this specific solution may not be unique, it is representative of the families of 
solutions that we identified using a variety of fitting strategies. The dotted line in 
Figure~\ref{fig2} illustrates our iterative fit, with the 95\% confidence interval shown 
as a gray shaded region. As expected from the suggestion of \cite{Cameron2017}, $\Rocrit$ 
is slightly above the solar value and establishes an approximate dividing line between 
stars with activity cycles and those with flat activity. Aside from the Sun, the cycling 
star with the highest value of Ro is 18\,Sco, while the stars above $\Rocrit$ all show 
flat activity. The flat activity stars $\rho$\,CrB and 16\,Cyg\,B have $\mathrm{Ro} < 
\Rocrit$, but the uncertainties extend well above $\Rocrit$. The magnetic grand minimum 
star HD\,166620 has $\mathrm{Ro} > \Rocrit$, but the uncertainty extends slightly below 
$\Rocrit$ where simulations suggest that grand minima occur \citep{Vashishth2023}.

The deviation of the estimated wind braking torque from the predictions of standard 
models can be traced to unexpected changes in both the large-scale magnetic field 
strength and the \xray luminosity. Our standard models predict \mbox{$B \sim P_{\rm 
phot}^{1/2}/\mathrm{Ro}$} (where $P_{\rm phot}$ is the photospheric pressure), and 
\mbox{$L_{\rm X} \sim L_{\rm bol}/\mathrm{Ro}^2$} \citep{vanSaders2013, Saunders2024}. In 
Figure~\ref{fig3} we show the ratio of the observations and the model predictions for our 
sample, raised to powers that reflect their relative contributions to the wind braking 
torque and normalized so that unity represents agreement. We see deviations of up to an 
order of magnitude in each of these observables as Ro approaches $\Rocrit$. The low blue 
point on the left side of Figure~\ref{fig3}a is $\iota$\,Hor, which is the only star in 
our sample with a wind braking torque that is not in the dipole-dominated regime of 
\cite{FinleyMatt2018}. Aside from this point, the trend with Ro resembles the pattern 
seen in Figure~\ref{fig2}. By contrast, the normalized ratios of $R'_{\rm HK}$ to $B_{\rm 
model}$ (gray points) are scattered around unity across the full range of Ro. This 
suggests that while our standard model correctly reproduces the evolution of the total 
magnetic field strength, it fails to predict the observed changes in the large-scale 
field (represented by $B_{\rm d}$) for stars near $\Rocrit$. As we have noted before, the 
solar $R'_{\rm HK}$ is dominated by a $\left<B\right>\sim 170$~G contribution from the 
unstructured quiet Sun, which dwarfs the $\sim 1$~G dipole component of the field 
inferred from ZDI maps \citep{Metcalfe2019}. The dipole field can disappear entirely with 
negligible impact on $R'_{\rm HK}$, but with severe consequences for the wind braking 
torque.

The change in the observed \xray luminosity shown in Figure~\ref{fig3}b may represent a 
decrease in the mass-loss rate that is captured by the empirical relation of 
\cite{Wood2018}. Three of the targets in our sample ($\epsilon$\,Eri, HD\,219134, 
$\tau$\,Cet) have direct inferences of the mass-loss rate from Ly$\alpha$ measurements 
(30, 0.5, 0.1~$\dot{M}_\odot$), which broadly agree with the values predicted from their 
\xray surface fluxes ($22.4^{+34.4}_{-16.3}, 0.31^{+0.31}_{-0.16}, 0.10^{+0.13}_{-0.08}\ 
\dot{M}_\odot$). This suggests that the observed decrease in wind braking torque is not 
an artifact of the empirical relation used to scale the \xray luminosity. Instead, it may 
reflect a genuine decrease in the mass-loss rate. Furthermore, the observed changes in 
$B_{\rm d}$ and $L_{\rm X}$ can both be understood as consequences of a near-critical 
dynamo and the resulting decrease in Poynting flux. As a star approaches $\Rocrit$, its 
internal dynamo becomes less efficient. This leads to a weaker large-scale magnetic field 
and a subsequent decrease in the outward-propagating Poynting flux, or magnetic energy 
flow. The resulting reduction in Poynting flux directly affects two key processes: it 
provides less energy for coronal heating, which lowers the \xray luminosity, and it 
reduces the magnetic pressure that accelerates the stellar wind, which in turn reduces 
the mass-loss rate.

In hindsight, a rotational threshold for the excitation of a global stellar dynamo is 
understandable considering the role of the Coriolis force, which imposes a tilt on 
emerging bipolar magnetic regions and imprints organizing flows on the convective 
patterns \citep{RolandBatty2025}. A smaller Joy's law tilt yields enhanced cancellation 
before the weaker differential rotation can separate the leading and trailing polarities. 
Inefficient meridional circulation then transports less residual magnetic flux to the 
polar regions, stunting the dipole field for the subsequent cycle. In addition, weaker 
differential rotation operating on a weaker dipole field is less efficient at winding up 
the poloidal field to produce toroidal flux tubes near the base of the convection zone, 
inhibiting flux emergence and leading to a downward spiral of the dipole field strength. 
The lower mass-loss rate appears to be a consequence of the diminishing magnetic energy 
that is available from the global dynamo. The increase in high-order magnetic complexity 
might also throttle the stellar wind, which escapes along open magnetic field lines 
\citep{Garraffo2015, Shoda2023}.

The Rossby number plays a central role in theoretical models of stellar winds, and it has 
been implicitly invoked in empirical studies. Our results strongly reinforce this 
hypothesis. Across a wide range of spectral types, Ro-scaled models predict reliable 
torques for the active stars. For less active stars, a dramatic decrease in torques is 
seen. WMB emerges at a consistent Ro across a wide range of stellar metallicities and 
convective overturn timescales. At higher Ro, we see unexpected decreases in both 
large-scale magnetic field strength and coronal heating. We caution that our observations 
do not reach the fully convective domain. We therefore cannot draw conclusions about the 
similarity of the dynamo mechanism in stars with radiative cores versus fully convective 
stars.

Future observations from ESA's PLATO mission could help expand the current sample of 
bright stars with measured rotation periods and stellar properties from asteroseismology 
\citep{Rauer2025}. Although a specific target list has not yet been released, the 
footprint of the first $49^\circ\!\times\!49^\circ$ field contains more than 100 bright 
stars with measured chromospheric activity levels \citep{Henry1996}. About half of these 
potential targets are in the low-activity range ($\log R'_{\rm HK} \la -4.9$) where the 
effects of WMB start to become apparent, and dozens are also bright enough for 
spectropolarimetry to be feasible with HARPSpol. The Galactic longitude of the field is 
well within the German half of the eROSITA all-sky survey \citep[eRASS;][]{Predehl2021}, 
so \xray surface fluxes might be available from eRASS DR2 (mid-2026) or ultimately DR3 
(late 2028). With patience and some luck, the current sample of bright solar-type stars 
that probe the WMB regime will be expanded substantially over the coming years, providing 
a broader context for our understanding of magnetic stellar evolution during the second 
half of main-sequence lifetimes.

\vspace*{12pt}\noindent 
The authors would like to thank Robert Cameron for discussions that substantially 
informed our interpretation of these observations, as well as Stefano Bellotti and Gaitee 
Hussain for sharing details about their published ZDI maps. Early support for this 
project came from the Max Planck Institute for Solar System Research (2018) and the 
Vanderbilt Initiative in Data-intensive Astrophysics (2021). T.S.M.\ and J.v.S.\ 
acknowledge support from NSF grants AST-2205919 and AST-2205888, respectively. O.K.\ 
acknowledges support from the Swedish Research Council (grant agreement 2023-03667) and 
from the Swedish National Space Agency. A.J.F.\ and V.S.\ received funding from the 
European Research Council (ERC) under the European Union's Horizon 2020 research and 
innovation programme (grant agreements 810218 WHOLESUN and 804752 CartographY, 
respectively).


\end{document}